# Localization Technologies for Indoor Human Tracking

Da Zhang, Feng Xia, Zhuo Yang, Lin Yao
School of Software
Dalian University of Technology
Dalian 116620, China
f.xia@ieee.org

Wenhong Zhao
College of Mechanical Engineering
Zhejiang University of Technology
Hangzhou 310032, China
zwh2010@sina.com

*Abstract*—The proliferation of wireless localization technologies provides a promising future for serving human beings in indoor scenarios. Their applications include real-time tracking, activity recognition, health care, navigation, emergence detection, and target-of-interest monitoring, among others. Additionally, indoor localization technologies address the inefficiency of GPS (Global Positioning System) inside buildings. Since people spend most of their time in indoor environments, indoor tracking service is in great public demand. Based on this observation, this paper aims to provide a better understanding of state-of-the-art technologies and stimulate new research efforts in this field. For these purposes, existing localization technologies that can be used for tracking individuals in indoor environments are reviewed, along with some further discussions.

## I. INTRODUCTION

In the past decades, wireless localization technologies have undergone considerable progress. They gradually play an important role in all aspects of people's daily lives [1], including e.g. living assistant, navigation, emergency detection, surveillance/tracking of target-of-interest, and many other location-based services. Reliable, accurate, and real-time indoor tracking services are required by people even more strongly than ever. For example, with the severely increasing number of elder people, the aging population has become a burning issue for all modern societies around the world. It has consequently become an urgent problem how to monitor those old people effectively when they are at home or inside other buildings [2]. In addition, more and more attention has been paid to context-aware applications which can make our life easier and convenient [3]. The realization of these applications is essentially based on location information.

For outdoor environments, GPS (Global Positioning System) plays a dominant role in localization [4]. However, it does not work well in indoor scenarios. This inefficiency is due to the weakness of signals emitted by GPS and their disability to penetrate most building materials. Therefore, GPS does not fit well in indoor environments where people spend most of their time. Even though GPS devices are becoming more and more promising and ponderable in the future and are able to provide sufficient precision for outdoor use, other effective technologies are demanded for indoor human/object tracking. To fulfill this requirement, various indoor localization technologies have been developed in the literature [5].

However, due to the complexity of indoor environments, the development of an indoor localization technique is always accompanied with a set of challenges, e.g. NLOS (none line of sight), multipath effect, and noise interference. These challenges result mainly from the influence of obstacles (e.g. walls, equipments, and human beings) on the propagation of electromagnetic waves. For instance, the mobility of people incurs changes in physical conditions of the environment, which might significantly affect the behavior of wireless radio propagation. Although these negative effects can not be eliminated completely, in recent years researches are constantly going on to improve the performance of indoor (human/object) tracking. There are several survey papers in the literature of indoor localization, e.g. [1,3,6]. To inspire new research efforts in the field, there is still a need of better understanding of state-of-the-art localization technologies. This paper is an attempt to serve for this purpose. We present a brief overview of existing localization techniques and methods, including signal measurement methods, positioning algorithms, networking techniques and systems, which can be used for indoor human tracking.

The rest of this paper is organized as follows. Section II will first illustrate basic concepts in indoor localization. The whole localization process is divided into two phases, i.e. signal measurement and position calculation. State-of-the-art localization methods and algorithms used in these two phases are reviewed in Sections III and IV respectively. In Section V, several popular network techniques used in the filed are discussed. Some well-known existing localization systems are also compared. Section VI concludes the paper with a discussion on open issues.

## II. PROBLEM STATEMENT

Indoor localization system, as Dempsey [7] defined, is a system that can determine the position of something or someone in a physical space such as in a hospital, a gymnasium, a school, etc. continuously and in real time. Based on this concept, consider a typical scenario of indoor human tracking. First, each reference sensor node (with known

This work is partially supported by Natural Science Foundation of China under Grant No. 60903153 and Zhejiang Provincial Natural Science Foundation of China under Grant No. Y108685.

position) sends a ranging request to the compatible mobile device attached to the target (i.e. people to be located). This device could be for example a cell phone or a PDA. Then the mobile device perceives the request signals and issues a ranging reply to the reference sensor. To this end, the sensor could calculate the transmission time between the sensor and the mobile device. Next, the sensor forwards the calculated time to a calculation center. Usually the calculation center could be a base station (BS) or a personal computer (PC). With powerful computational capability, the calculation center processes the received data using some positioning algorithm to obtain the position of the target.

From the above description, we can see that in order to obtain the physical position of the target-of-interest in indoor environments, two steps are usually needed [3,8]: first, some position-related signal parameters corresponding to wireless communications between the target and the sensor are measured; and then, the physical position of the target is calculated based on these signal parameters.

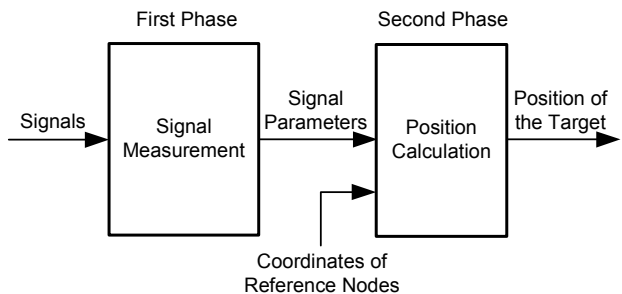

Figure 1. Two phases in localization

As shown in Fig. 1, the whole localization process can generally be divided into two phases: signal measurement and position calculation. In the first phase, some signals are transmitted between the target node (representing the communication entity attached to the people) and a number of reference (sensor) nodes. During this process, some properties of these signals, such as arrival time, signal strength, and direction, are captured by the receivers. As such, certain signal parameters, such as TOA (Time of Arrival), TDOA (Time difference of Arrival), RSS (Received Signal Strength), and AOA (Angle of Arrival), will be extracted. Various methods used in this phase will be covered in the next section.

In the second phase, the physical position of the target node will be determined based on the signal parameters obtained in the first phase. The most common technique used here is based on ranging, whereby distance or angle approximations are obtained [3]. In this context, geometric approaches will be employed to calculate the position of the target node as the intersection of position lines obtained from the position-related parameters at reference nodes. Trilateration and triangulation are two most popular geometric approaches, which will be introduced in Section IV. In addition, since signal measurements in real systems are only accurate to some extent (especially in indoor environments), optimization-based statistical techniques are often used to filter measurement noise and improve accuracy of the result.

## III. SIGNAL MEASUREMENT

In this section, we elaborate on various measurement methods involved in the first phase of localization (see Fig. 1). More specifically, our focus is on the three most popular categories of methods for this phase: one category is time based methods; another is the angle based method (i.e. AOA); the third is the received signal strength based method (i.e. RSS). In the following, we describe relevant technologies belonging to each of these categories, with some related work reviewed.

### A. Time-based Methods

#### 1) Time-of-Arrival (TOA)

With TOA, the distance between the transmitting node and the receiving node is deduced from the transmission time delay and the corresponding speed of signal as follows:

$$R = time \times speed \quad (1)$$

where *speed* denotes the traveling speed of the signal, *time* the amount of time spent by the signal travelling from the transmitting to the receiving node, and $R$ the distance between the transmitting node and the receiving node. Since *speed* can be regarded as a known constant, $R$ can be computed by observing *time*.

One of the most widespread techniques uses TOA jointly with UWB technology to achieve higher precision [9]. An overview of this combination has been given in [10]. It has been recognized that TOA technique can best deal with fine time resolution with the help of UWB technology [11]. While TOA technique has a restrict requirement for synchronization, this inefficiency can be compensated by UWB due to its nature of sensitivity to time. UWB technology uses short pulse duration to filter out the signals caused by reflection to improve the overall performance. As a consequence, the combination of these two technologies is predominant in the indoor localization field at present.

#### 2) Time Difference-of-Arrival (TDOA)

This technology uses two different kinds of transmitting signals. The time difference between these two kinds of signals is used to reconstruct the transmitting node's position. The calculation is based on the following equation:

$$\frac{R}{c_1} - \frac{R}{c_2} = t_1 - t_2 \quad (2)$$

In (2), $c_1$ denotes the speed of one kind of signal, $c_2$ the speed of another kind of signal, $t_1$ and $t_2$ the time for these two signals travelling from one node to the other respectively, and $R$ still the distance between the transmitting node and the receiving node.

A considerable number of works have explored TDOA-based methods. For instance, Takabayashi *et al*. [12] employ TDOA technique to realize target tracking. This technology is based on EKF (Extended Kalman Filter), FDOA (Frequency Difference of Arrival) and TDOA technologies. Unlike conventional methods which require enough sensors to estimate the position of target, the authors only use utilizable sensors to calculate the position. Therefore, the approach is pretty suitable in environments where the number of sensors is not sufficient.

*3) Round Trip Time (RTT)*

This measurement method emerges with the goal of solving the problem of synchronization incurred by TOA. With RTT, the distance is calculated as follows:

$$R = \frac{(t_{RT} - \Delta t) \times speed}{2} \quad (3)$$

where $t_{RT}$ denotes the amount of time needed for a signal to travel from one node to the other and back again, $\Delta t$ the predetermined time delay required by the hardware device to operate at the receiving node, and *speed* the speed of the transmitting signal. It is clear that RTT is a reciprocal technology [13,14]. Instead of using two local clocks in both nodes to calculate the delay (as TOA technology does), it uses only one node to record the transmitting and arrival time. Therefore, to some extent, this technology solves the problem of synchronization.

Although time-based measurement methods are now in widespread use, they are limited by restrict requirement of synchronization [15]. That means it is necessary to set synchronized clocks to both the transmitting nodes and the receiving nodes. A consequence is that it would be very costly in order to install the system and maintain the accuracy at runtime. The RTT based techniques are able to solve the problem of synchronization to a certain extent. However, they increase the complexity of the whole (reference) sensor system to $O(n^2)$. In a sensor system consisting of *n* nodes, it takes every node *n* times to locate its position through message exchanging. Additionally, with the RTT technology, other uncertain factors (e.g. noise) coexist during the process of time measurement. Therefore, the problem of synchronization deserves further investigation.

*B. Angle-of-Arrival (AOA)*

With respect to AOA-based techniques [5,6,8,16], the reference nodes or the target node has the capability of measuring the angle of arrival based on information obtained. For this purpose, techniques like angle diversity may be utilized in order to exploit the directionality of the receiver. Usually, direction finding can be accomplished by either with directional antennas or with an array of antennas. The main principle behind the AOA measurement via antenna arrays consists in that differences in arrival times of an incoming signal at different antenna elements include the angle information given that the array geometry is known. With AOA, no time synchronization between nodes is required.

AOA-based techniques have been widely used in the literature. For example, Yang *et al.* [17] combine the AOA technology with another TOA-based technique, i.e. TDOA, to achieve higher accuracy. In [18], based on the basic AOA concept, the authors develop a localization technique using cooperative AOA approach. Instead of requiring sets of acoustic model arrays and antenna arrays in each node like other conventional AOA based techniques, the approach only needs one set of acoustic model array and antenna array in each node by introducing the concept of super nodes which are actually virtual AOA-capable nodes.

Unsurprisingly, AOA-based techniques have their limitations. Since AOA-based methods are highly sensitive to multi-path and NLOS, it is not suitable for indoor localization sometimes. As the distance increases, the localization precision will decrease. In addition, technologies based on AOA require additional antennas with the capacity to measure the angles. This increases the cost of the whole system.

*C. Received Signal Strength (RSS)*

For the RSS based techniques, the distance is measured based on the attenuation introduced by the propagation of the signal from the transmitting node to the receiving node. An empirical mathematical model to calculate the distance according to signal propagation is as follows [19,20]:

$$p(R) = p(R_0) - 10n\log(\frac{R}{R_0}) - \begin{cases} nW \times WAF & (nW < C) \\ C \times WAF & (nW \geq C) \end{cases} \quad (4)$$

The attenuation formula can be expressed in (4), where $R$ denotes the distance between the transmitter and the receiver, $R_0$ a reference distance, $p(R)$ and $p(R_0)$ the signal strength received at $R$ and $R_0$ respectively, $nW$ the number of obstacles between the transmitter and the receiver, $WAF$ the attenuation factor of the wall, $C$ the maximum number of obstacles between the transmitter and the receiver, and $n$ the routing attenuation factor which could be determined by both theoretical and empirical calculations.

Based on the RSS technology, several methods have been proposed to estimate the position of the target-of-interest. For example, the fingerprint-based solution [21] for target positioning is the most typical application of RSS technology. In general, we can divide the fingerprint methodology into two steps: sampling (offline) and matching (online). In the sampling step, a database is created offline to store the radio signal map consisting of the geographical positions and the corresponding signal strengths. These signals may be e.g. sound, light, color, and human movement, among others. In the matching step, the relevant signals collected for the target (node) are compared against the pre-stored records of the geographic-signal map. By doing so, it will be able to determine where the target is, as long as any record in the database is matched.

IV. POSITION CALCULATION

Based on the signal parameters measured in the first phase and the known coordinates of reference nodes, it is then possible to calculate the physical position (i.e. coordinates) of the target in the second phase of localization (Fig. 1). To do this, the trilateration and triangulation techniques are commonly used. In addition, statistical techniques could be employed to improve the solution accuracy by coping with measurement noise. In this regard, we will introduce a very popular parametric approach: maximum likelihood estimation, though there are many other approaches in the literature.

## A. Trilateration

As illustrated in Fig. 2, the trilateration based positioning algorithm uses three fixed non-collinear reference nodes to calculate the physical position of a target node (in 2D).

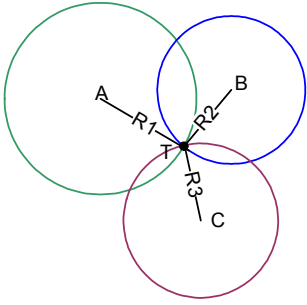

Figure 2. Trilateration-based positioning

Based on the coordinates of three reference nodes: $A(x_1, y_1)$, $B(x_2, y_2)$, and $C(x_3, y_3)$, and the corresponding distances from each reference node to the target node: $R_1$, $R_2$, and $R_3$, we can obtain the following equations:

$$\begin{cases} (x_1 - x)^2 + (y_1 - y)^2 = R_1^2 \\ (x_2 - x)^2 + (y_2 - y)^2 = R_2^2 \\ (x_3 - x)^2 + (y_3 - y)^2 = R_3^2 \end{cases} \quad (5)$$

where $(x, y)$ denotes the (unknown) coordinates of the target $T$.

Based on the trilateration algorithm, Han *et al.* [22] further improve localization performance by taking into account the layout of the three reference nodes. The work has approved that the trilateration algorithm can best demonstrate its advantages when the three reference nodes are deployed in the vertices of equilateral triangles. Yang and Liu [23] consider the effect of noisy environments, and use different confidence coefficients for three nodes to guarantee the quality of trilateration.

## B. Triangulation

When AOA measurements are available, triangulation can be used to determine the position of the target node. Instead of measuring distances between nodes as trilateration does, triangulation-based positioning is based on the measurement of angles, though they work in a similar manner. In most situations, triangulation can be transformed to trilateration since the distance between nodes can be reconstructed from the bearings between them. However, compared to trilateration, only two reference nodes are needed for triangulation (in 2D), instead of three.

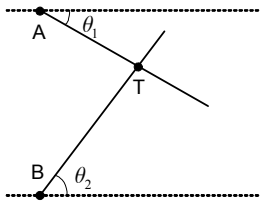

Figure 3. Triangulation-based positioning

With triangulation, the position of the target node can be determined by the intersection of several pairs of angle direction lines [6]. As shown in Fig. 3 where $A$ and $B$ represent reference nodes, after obtaining the angles $\theta_1$, and $\theta_2$, the physical position of $T$ (representing the target to be located) could then be calculated based on the predetermined coordinates of the reference nodes.

## C. Maximum Likelihood Estimation (MLE)

MLE is a popular statistical method used for addressing the problem of measurement uncertainty in localization. In this subsection, we will describe MLE in the context of trilateration-based positioning. Suppose the MLE method [20] uses $n$ reference nodes to calculate the target node's coordinates (generally $n \geq 3$). The relevant equations are given below:

$$\begin{cases} (x_1 - x)^2 + (y_1 - y)^2 = R_1^2 \\ (x_2 - x)^2 + (y_2 - y)^2 = R_2^2 \\ \vdots \\ (x_n - x)^2 + (y_n - y)^2 = R_n^2 \end{cases} \quad (6)$$

In (6), using every equation to subtract the subsequent one, we will get:

$$\begin{cases} x_1^2 - x_n^2 - 2(x_1 - x_n)x + y_1^2 - y_n^2 - 2(y_1 - y_n)y = R_1^2 - R_n^2 \\ \vdots \\ x_{n-1}^2 - x_n^2 - 2(x_{n-1} - x_n)x + y_{n-1}^2 - y_n^2 - 2(y_{n-1} - y_n)y = R_{n-1}^2 - R_n^2 \end{cases}$$

Let $A = \begin{bmatrix} 2(x_1 - x_n) & 2(y_1 - y_n) \\ \vdots & \vdots \\ 2(x_{n-1} - x_n) & 2(y_{n-1} - y_n) \end{bmatrix}$,

$b = \begin{bmatrix} x_1^2 - x_n^2 + y_1^2 - y_n^2 + R_n^2 - R_1^2 \\ \vdots \\ x_{n-1}^2 - x_n^2 + y_{n-1}^2 - y_n^2 + R_{n-1}^2 - R_n^2 \end{bmatrix}$, and $X = \begin{bmatrix} x \\ y \end{bmatrix}$,

then we have $AX = b$.

By adopting the minimum variance estimation method, the coordinates $(x, y)$ of the target can be calculated based on the following equation:

$$X = (A^T A)^{-1} A^T b \quad (7)$$

Besides estimating the coordinates of the target, the authors of [24] and [25] also use MLE to solve the problem of synchronization by predicting the uncertain parameters in time bias. Particularly, Tian *et al.* [24] thoroughly analyzed the source factors causing time bias in different transmission stages. This analysis contributes to further research on resolving different aspects of the synchronization problem.

## V. NETWORKING TECHNIQUES AND SYSTEMS

In this section, we first outline the signal technologies that are commonly used, discussing their pros and cons; then classify existing systems into several groups and make a comparison amongst them. It is worthy to note that not only the measurement method and positioning algorithm but also the signal technology of a localization system can have a heavy impact on the accuracy of localization.

TABLE I. LOCALIZATION SYSTEMS

| System | Network | Accuracy | Method | Overall Evaluation (A: Advantage; D: Disadvantage) |
|---|---|---|---|---|
| WhereNet [31] | RFID | 2m to 3m | TDOA | A: Uniquely identify equipment and person. <br> D: Need numerous infrastructure components |
| RADAR [32] | WLAN | 2.26m out of 312m$^2$ | Triangulation | A: Reuse the existing WLAN infrastructure. <br> D: Low level accuracy, no consideration of privacy |
| EKAHAU [33] | WLAN | 1m | RSSI | A: Low cost and power level of the battery. <br> D: Low level accuracy and only provide 2D location information. |
| COMPASS [34] | WLAN | 1.65m out of 312m$^2$ | Fingerprint | A: Consider the orientation impact of the user. <br> D: Only consider single user. |
| Ubisense [35] | UWB | Tens of centimeters | TDOA and AOA | A: No requirement of line-of-sight; large coverage area; 3D location; high accuracy <br> D: The price of the system is high. |
| Active Badge [26] | Infrared | Room level | RSS | A: Address privacy <br> D: Low accuracy; long transmission period; influenced by fluorescent light and sunlight |
| Firefly [27] | Infrared | 3.0mm | Not available | A: High level accuracy; small measurement delay of 3 ms <br> D: Use wire to connect tags and the coverage area is limited to 7m. |
| OPTOTRAK [28] | Infrared | 0.1mm to 0.5mm | Not available | A: High accuracy; able to measure relative motions on the different parts of one object. <br> D: Limited by line-of-sight requirement. |
| Sonitor [36] | Ultrasound | Room level | Not available | A: Energy efficient <br> D: Low level accuracy |
| IRIS_LPS [37] | Infrared | 16cm out of 100m$^2$ | Triangulation | A: Larger covered area <br> D: Subject to interference from florescent light and sunlight |
| Active Bat [29] | Ultrasound | 3cm out of 1000m$^2$ | Multilateration | A: Cover large area; provide 3-D position. <br> D: Subject to reflections of obstacles; use a large number of transmitters on the ceiling. |
| Cricket [30] | Ultrasound, RF | 10cm | TOA and triangulation | A: Address privacy; low cost, decentralized administration. <br> D: More energy consumption |

## A. Infrared (IR) Based Systems

The most prominent advantage of IR is its wide availability since many devices are equipped with IR sources, such as mobile phones, TV, printer, PDAs, and so forth. In addition, since the whole infrastructure is very simple, it does not need costly installation and maintenance. However, due to its requirement of line-of-sight and its inability to penetrate opaque obstacles, it can not be applied to some kinds of indoor scenarios in which the environment is pretty complex. Besides, it is subject to interference of other sources of IR devices. Several systems are based on this technology, including Active Badge [26], Firefly [27], and OPTOTRAK [28], for example.

## B. Radio Frequency (RF) Based Systems

Systems designed based on RF can cover larger distance since it uses electromagnetic transmission, which is able to penetrate opaque objects such as people and walls. Besides, a RF system can uniquely identify people or objects tracked in the system. In the literature, triangulation and fingerprint techniques are widespread used in RF based systems. Based on this technology, RFID (Radio Frequency Identification), WLAN (Wireless Local Area Network), Bluetooth, wireless sensor networks, UWB (Ultra Wide Band) are created. In addition, RF based technologies are further divided into narrow band based technologies (RFID, Bluetooth and WLAN) and wide band based technologies (UWB). Amongst these technologies, UWB is the most accurate and fault-tolerant system that has a widespread usage in indoor localization.

## C. Ultrasound Based Systems

Although the systems based on ultrasound technology is relatively cheap, the precision is lower in comparison with IR-based systems due to the reflect influence. Additionally, this kind of systems is always associated with RF technology to fulfill the synchronization requirement, which may increase the cost of the whole system. Active Bat [29] and Cricket [30] are example applications of ultrasound technology.

In Table I [1,5,6], we make a comparison between some major localization systems in various aspects, including accuracy, advantages and disadvantages, networking technologies and localization methods.

## VI. CONCLUSION

In this paper, we presented a brief overview of state-of-the-art localization technologies for tracking individuals in indoor environments. Some related works are reviewed. Despite the great progress made in recent years, there are a number of open issues that need to be addressed. Examples include e.g. continuously tracking people travelling between indoor and outdoor environments, solving synchronization problems, reducing the impact of noise interference, and improving energy efficiency. Although some previous technologies are concerned with these issues, they might suffer from various limitations, e.g. increase in the cost of the whole system, precision deficiency, and severe computational overhead. Innovative research efforts are expected to tackle these issues in the near future.